# The Relevance of Assumptions and Context Factors for the Integration of Inspections and Testing

Frank Elberzhager, Robert Eschbach
Fraunhofer IESE
Kaiserslautern, Germany
{frank.elberzhager,
robert.eschbach}@iese.fraunhofer.de

Jürgen Münch
University of Helsinki
Helsinki, Finland
juergen.muench@cs.helsinki.fi

*Abstract*—Integrating inspection processes with testing processes promises to deliver several benefits, including reduced effort for quality assurance or higher defect detection rates. Systematic integration of these processes requires knowledge regarding the relationships between these processes, especially regarding the relationship between inspection defects and test defects. Such knowledge is typically context-dependent and needs to be gained analytically or empirically. If such kind of knowledge is not available, assumptions need to be made for a specific context. This article describes the relevance of assumptions and context factors for integrating inspection and testing processes and provides mechanisms for deriving assumptions in a systematic manner.

## I. INTRODUCTION

During the development and maintenance lifecycle of software-intensive systems and services, typically many processes are followed, such as inspection or testing processes. These processes capture the application of one or more techniques such as code reading or structural testing. Although the overall lifecycle model typically links all of these processes, many processes are performed separately without deeper integration. Systematic integration of different quality assurance processes, for instance, is often missing, although it promises to deliver certain benefits, such as higher efficiency and effectiveness regarding quality assurance, and, as a consequence, reduced overall costs.

In order to exploit these benefits, knowledge regarding the relationships between different processes is required. Such relationships are usually context-specific and not generally applicable. Therefore, it is necessary to check whether reliable evidence regarding such relationships exists in a given context (e.g., stored in an experience base [13]). If such evidence does not exist, assumptions need to be made regarding relationships between the processes to be considered. An example assumption might be that the distribution of defects found regarding certain defect types is similar for inspection and testing for the same artifact. Therefore, it might be beneficial to use the defect distribution from inspections for creating the test cases. Assumptions that describe certain relationships can initially be taken from the literature or from different contexts, but need to be analyzed with respect to their validity in the given context. Evidence regarding defined assumptions can be gathered in different ways (e.g., analytically, empirically), and has to be continuously reevaluated and updated due to the fact that context factors can change and thus, assumptions initially defined and proven to be correct can become wrong.

One type of process integration is the integration of inspection and testing processes. Such integration promises several benefits, for instance improved defect detection effectiveness or reduced overall quality assurance costs. However, the benefits achieved depend on knowledge regarding the relationships between inspection and testing processes, especially knowledge regarding the distribution of defects in inspections and testing. If such knowledge is available, it can be used to balance and focus inspection and testing activities.

For instance, using the assumption stated before that both quality assurance activities mainly find defects of the same defect types, testing activities may be focused on those defect types that inspection has primarily found before. Or consider the assumption of a Pareto distribution for defects found; then testing activities may be focused on those parts where inspection has found most of the defects before.

There exist several models that describe how assumptions can be identified and evaluated. Jeffery and Scott [1], for instance, developed a model for scientific inquiry, starting by observing a phenomenon in the real world, understanding it, and developing a theory that explains the observed phenomenon. Such a theory has to be validated and refined by means of theory testing, replication, theory revision, and reevaluation. Jeffery and Scott use two concrete examples, i.e., 'software cost modeling and estimation' and 'software inspections', in order to demonstrate which of these steps were conducted and what the implications are. While for the first example, valid theories could be derived and demonstrated, this could not be done for the second example. The authors state that there exists "confusion in the empirical inspection literature", which "is a result of insufficient expression of theory, a consequent lack of models, and too little attention in the experiments to the justification for the hypotheses under test" [1]. Moreover, Bertolino [11] states that for testing, no universal theory exists either. Sjoberg et al. [12] conclude that almost no software engineering specific theories are reported in the literature.

From the viewpoint of the authors, instead of finding a theory first, in many cases it seems to be more promising to get context-specific evidence first. Later on, a valid theory might be derived.

This article is structured as follows: Section 2 demonstrates how to derive and evaluate assumptions that describe relationships between inspection and test processes. Context-specific relationships between inspection and testing defects are presented in Section 3. Finally, Section 4 summarizes and concludes this article. An extended version of this article includes related work and an exemplary application of the concepts [14].

## II. DERIVATION AND EVALUATION OF CONTEXT-SPECIFIC RELATIONSHIPS

The field of empirical software engineering presents various concepts that guide the way from initial observations to evaluated theories. One main objective is to improve the understanding regarding processes, products, and resources, and to build up solid knowledge in order to be able to predict future situations and make them more controllable.

In contrast to the model by Jeffery and Scott [1] mentioned in the introduction a more detailed model is proposed by Endres and Rombach [2]. The model starts with observations, which may be facts or impressions regarding certain relationships in a given context. When an observation reappears, one can take advantage of it. Repeatable observations are often defined as so-called laws. The authors define a law as "a statement of an order or relation of phenomena that, so far as is known, is constant under certain conditions". Exemplary laws in the field of quality assurance mentioned by the authors are that a developer is unable to test his own code or that about 80 percent of the defects come from 20 percent of the modules. Laws are explicitly derived based on repeatable observations and lessons learned from different contexts. Because laws are based on strong empirical evidence, they can be seen as generalized observations that explain how things happen, independently of a concrete environment (though some situations may exist where a law might be wrong). Furthermore, future observations can be predicted based on laws.

A law can be explained by a theory. "A theory is a deliberate simplification of factual relationships that attempts to explain how these relationships work" [3]. Sjoberg et al. [12] state that "in mature sciences, building theories is the principal method of acquiring and accumulating knowledge that may be used in a wide range of setting." Therefore, if a law is found out, the next step is to find explanations for the observations, which shifts the level of understanding towards a theory. A theory itself can then be confirmed by future observations (until it may be rejected due to new insights and knowledge that falsifies the theory).

Fig. 1 summarizes the concepts as stated by Endres and Rombach [2]. Finally, besides laws, the authors introduced two additional constructs in order to be able to describe relationships that are currently not grounded on strong empirical evidence. A hypothesis is a statement that is only tentatively accepted, for example, only in a certain context. Additional evidence is needed in order for a hypothesis to become a law. A conjecture describes the lowest level in this hierarchy and is a guess or belief only.

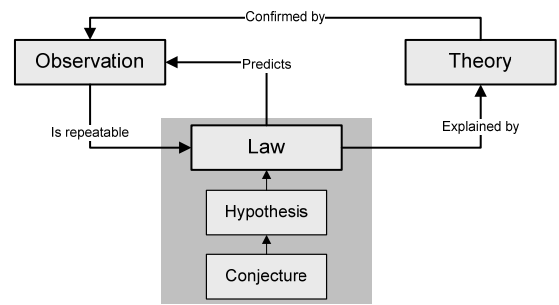

Figure 1. Concepts of empirical software and systems engineering according to Endres and Rombach [2]

Endres and Rombach [2] describe three stringent criteria for accepting existing knowledge as a law: First, an underlying hypothesis exists that has been validated; second, the explicit kinds of studies used for the evaluations are determined (e.g., case study, experiment); and third, replications of studies are conducted in different environments. However, it is sometimes more difficult to distinguish hypotheses and conjectures. Carver et al. [10] or Bertolino [11], for instance, utilize the term assumptions when referring to empirical studies.

Consequently, an adaptation of the model proposed by Endres and Rombach [2] is performed, and a distinction into assumptions and evaluated assumptions (i.e., evidence) is made in the following. An assumption describes context-specific relationships that are observed or seem to be useful, but are not empirically grounded. In contrast, evaluated assumptions are based on empirically valid results that are accepted in the given context. In order to explain the evaluated assumptions, the results can be used to derive a theory for the given context.

Instead of starting with observations in order to derive assumptions, sometimes a theory is stated first, which subsequently has to be confirmed or rejected based on assumptions derived from the theory. Fig. 2 summarizes these concepts.

Context-specific relationships can be derived analytically or empirically.

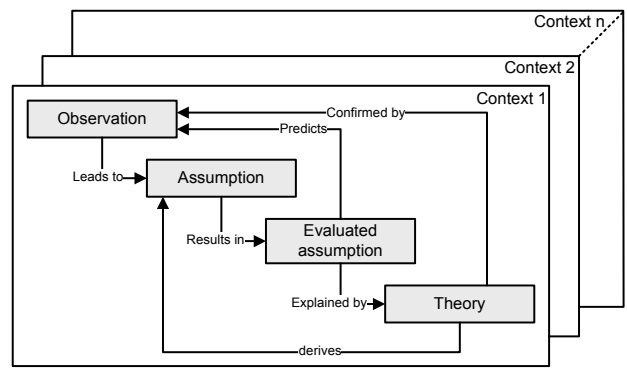

Figure 2. Concepts for empirical software engineering

*1. Analytically:* Based on a systematic analysis of processes, product structures, and context characteristics, assumptions regarding relationships can be derived for a specific context in a logical manner. Example assumptions could be: (a) Certain defect types cannot be found (or not sufficiently found) with inspections – therefore, testing activities need to be adjusted accordingly; (b) only parts of the system can be inspected due to reasons such as late or dynamic integration – therefore, testing needs to be especially focused on these parts.

*2. Empirically:* Assumptions regarding relationships can be derived based on (a) general empirical observations and knowledge, and (b) new experiences and observations from the given context. First, empirical knowledge from different contexts can be used and adapted to the given context. One example assumption is: If a significant number of inspection defects are found in a certain module, it is expected that more defects are to be found in this module during testing. Such a Pareto distribution is shown in general for defects in different contexts and describes an evaluated assumption (Endres and Rombach [2] stated a law regarding this kind of defect distribution). This assumption has to be checked and evaluated in each new context. Second, when such assumptions are evaluated empirically in a given context, new observations can be made, resulting in adapted or new assumptions, i.e., new empirical knowledge about relationships is gained.

The result of an evaluation of an assumption can be positive or negative. If the assumption was confirmed, all relevant context factors and the results should be packaged and additional evaluations should be performed in order to increase the significance of the evidence (i.e., the empirical evidence regarding the assumption). If the assumption was not confirmed, this might have different reasons: First, the assumption may be wrong in general. Then an alternative assumption has to be stated. Second, context factors were not considered or behaved differently in the given context. Consequently, these factors should be considered in the future and the assumption has to be adapted. Third, the assumption did not achieve the desired level of significance. Thus, slight adaptations of the assumption might be conceivable and additional evaluations are necessary.

Beside an initial evaluation of assumptions in order to understand certain relationships, continuous evaluations are necessary to improve the observed phenomenon in the best possible way and to enable further adaptations, for example, due to subsequent context changes.

A comprehensive evaluation of assumptions, both analytically and empirically derived ones, may result a profound basis of empirical evidence. Finally, this might result in new theories.

### III. CONTEXT-SPECIFIC RELATIONSHIPS BETWEEN INSPECTION AND TEST DEFECTS

Jeffery and Scott [1] stated that a profound underlying theory in the area of software inspections is missing. This lack is even more critical when inspection and testing techniques are combined in order to exploit certain synergy effects, such as reduced effort or higher defect detection rates. Consequently, there is no way to avoid determining assumptions regarding relationships that have to be analyzed afterwards in a systematic manner. However, there exist a number of accepted evaluated assumptions or laws, as Endres and Rombach [2] call them, which can be used and adapted to the area of combined quality assurance techniques. Due to unknown or partially unknown relationships, different exemplary assumptions are mentioned in the following that may form a starting point for evaluating them and that might lead to theories in the future. A distinction is made between analytically and empirically derived assumptions. A short explanation of each assumption is given next. Each of these assumptions has to be evaluated in relevant contexts in the future in order to show whether they are true or wrong.

#### A. Analytically Derived Assumptions

Various assumptions can be derived analytically, i.e., they can be determined logically. Some arbitrary examples are presented next.

> *Assumption A1:* If no defined selection criterion is used to determine parts of a system that should be inspected, it is expected that a significant number of defects still remain in those parts that are not inspected. Consequently, testing should be focused especially on those uninspected parts of a system.

Sometimes, inspections of certain parts are skipped due to external reasons that are not related to quality assurance (e.g., time constraints, missing resources). This may lead to re-planning of quality assurance activities. Consequently, a testing activity should be focused on the remaining parts of the system to find additional defects.

> *Assumption A2:* Inspection and testing activities find defects of various defect types with different effectiveness. For inspections, this includes, e.g., maintainability problems. For testing, this includes, e.g., performance problems. Consequently, inspection and testing activities should be focused on those defect types that are most convenient to find.

Among others, Gilb and Graham [4] already mentioned that inspection and testing complement each other. This also means that they are able to find different kinds of defects. For example, Mantyla and Lassenius [5] mention that code inspections find evolvability defects (e.g., defects affecting documentation or structure) that cannot be found by testing activities. One reason is that those maintainability problems do not affect functionality that is tested later. In contrast, problems that are only found when the system is running, such as performance problems, can be found better or only with testing. However, despite such defect types that are easy to assign to one quality assurance technique, it is unclear for many other defect types whether they can be found better with inspections or with testing.

#### B. Empirically Derived Assumption

As mentioned above, little empirical evidence exists in the area of combined inspection and testing techniques. Therefore, empirical evidence from related areas is taken and adapted as a starting point.

> *Assumption E1:* Parts of the system where a large number of inspection defects are found (i.e., a Pareto distribution of defects is observed) indicate more defects to be found with testing.

A large number of different studies performed in various environments showed that an accumulation of defects can be observed rather than an equal distribution of defects. A lot of empirical evidence exists that shows such a Pareto distribution of defects, i.e., about 80 percent of the defects are often found in about 20 percent of the modules [6]. Some recent studies have confirmed these results [7][8].

> *Assumption E2:* Parts of the system where a large number of inspection defects are found (i.e., a Pareto distribution of defects is observed) and which are of small size indicate more defects to be found with testing.

A size metric is often used to prioritize defect-prone parts and thus, to focus a testing activity. Though this metric is often applied, a number of studies showed inconsistent results when size is applied as the sole metric for predicting defect-prone modules. Thus, a combination of assumption E1 (i.e., Pareto distribution of defects) and size might lead (a) to a more detailed and fine-grained assumption and (b) to a better prediction of defect-prone parts than using a size metric alone.

> *Assumption E3:* Defects of the defect types that are found most often by inspections (i.e., a Pareto distribution of defects of certain defect types is observed) indicate more defects of the defect types to be found with testing.

An accumulation of defects of certain defect types can also be observed in several studies rather than an equal distribution of defect types, independent of a concrete defect classification [9]. Ohlsson et al. [14] stated that the majority of quality costs are often caused by very few defect types. However, one has to be aware that this is not necessarily so for each defect type.

In conclusion, various assumptions are possible when analyzing relationships between inspection and testing techniques. Some of them seem to be contradictory, such as assumptions A2 and E3; in this case, future evaluations might show which direction is true in certain contexts. The defined assumptions can serve as a starting point for such evaluations.

## IV. Summary and Conclusion

This article emphasized the relevance of defining assumptions and considering context factors when relationships between certain software development processes are not well understood or not known. From our point of view, this is a major contribution to systematically analyzing relationships and gathering solid empirical evidence that can lead to evaluated assumptions and theories to explain the observations made. Furthermore, such knowledge can be used to improve and control processes. Sjoberg et al. [12] identified five challenges with respect to empirical studies: more empirical studies, increased quality and relevance of empirical studies, synthesizing evidence, and theory building. Moreover, they emphasized the importance of performing empirical studies in order to understand observations and to be able to derive well-grounded theories based on evaluated assumptions.

Due to the lack of solid theories in the field of software inspections, it is essential to substantiate research that combines inspection and testing processes by systematically defining assumptions and evaluating them, and by considering context factors.

Regarding future work, one main step is to gather more empirical evidence, which includes a detailed analysis of various assumptions in different contexts.

### Acknowledgment

This work has been funded by the Stiftung Rheinland-Pfalz für Innovation project "Qualitäts-KIT" (grant: 925). We would like to thank Sonnhild Namingha for proofreading.